\documentclass[aps,prd,twocolumn,superscriptaddress,showpacs,nofootinbib]{revtex4-2}

\usepackage[dvipsnames]{xcolor}
 
\usepackage{booktabs, cellspace}
\usepackage{siunitx}
\usepackage{hyperref}

\newcommand*{\specialcellbold}[2][b]{%
  \bfseries
  \sisetup{text-rm = \bfseries}%
  \begin{tabular}[#1]{@{}c@{}}#2\end{tabular}%
}

\usepackage{graphicx}
\usepackage[caption=false]{subfig}
\usepackage{amssymb}
\newcommand{\Msun}{\ensuremath{\mathrm{M}_\odot}}

\begin{document}


\title{A wider look at the gravitational-wave transients from GWTC-1 using an  unmodeled reconstruction method}

\author{F.~Salemi}
\affiliation {Albert-Einstein-Institut, Max-Planck-Institut f\"ur Gravi\-ta\-tions\-physik, D-30167 Hannover, Germany }
\author{E.~Milotti}
\affiliation {Dipartimento di Fisica, Universit\`a di Trieste and INFN Sezione di Trieste, Via Valerio, 2, I-34127 Trieste, Italy } 
\author{G.~A.~Prodi}
\affiliation {Universit\`a di Trento, Dipartimento di Fisica, I-38123 Povo, Trento, Italy }
\affiliation {INFN, Trento Institute for Fundamental Physics and Applications, I-38123 Povo, Trento, Italy }
\author{G.~Vedovato}
\affiliation {INFN, Sezione di Padova, I-35131 Padova, Italy }
\author{\\ C.~Lazzaro}
\affiliation {Gran Sasso Science Institute, Via F. Crispi 7, I-67100, L'Aquila, Italy}
\author{S.~Tiwari}
\affiliation {Physik-Institut, University of Zurich, Winterthurerstrasse 190, 8057 Zurich, Switzerland }
\author{S.~Vinciguerra}
\affiliation {Albert-Einstein-Institut, Max-Planck-Institut f\"ur Gravi\-ta\-tions\-physik, D-30167 Hannover, Germany }
\author{ \\ M.~Drago}
\affiliation {Gran Sasso Science Institute, Via F. Crispi 7, I-67100, L'Aquila, Italy}
\affiliation {INFN, Laboratori Nazionali del Gran Sasso, I-67100 Assergi, Italy}
\author{S.~Klimenko}
\affiliation {University of Florida, Gainesville, FL 32611, USA }


\date{\today}

\begin{abstract}
In this paper, we investigate the morphology of the events from the GWTC-1 catalog of compact binary coalescences as reconstructed by a method based on coherent excess power: we use an open-source version of the coherent WaveBurst (cWB) analysis pipeline, which does not make use of waveform models. 
The coherent response of the LIGO-Virgo network of detectors is estimated by using loose bounds on the duration and bandwidth of the signal. 
This pipeline version reproduces the  same results that are reported for cWB in recent publications by the LIGO and Virgo collaborations.
In particular, the sky localization and waveform reconstruction are in a good agreement with those produced by methods which exploit the detailed theoretical knowledge of the expected waveform for compact binary coalescences. However, 
in some cases cWB also detects features in excess in well-localized regions of the time-frequency plane.
Here we focus on such deviations and present the methods devised to assess their significance.
Out of the eleven events reported in the GWTC-1, in two cases -- GW151012 and GW151226 -- cWB detects an excess
of coherent energy after the merger ($\Delta t \simeq 0.2$ s and $\simeq 0.1$ s, respectively) with p-values that call for further investigations ($0.004$ and $0.03$, respectively), though they are not  sufficient to exclude noise fluctuations.
We discuss the morphological properties and plausible interpretations of these features. 
We believe that the methodology described in the paper shall be useful in future searches for compact binary coalescences.
\end{abstract}


\maketitle

\section{Introduction \label{intro}}
Gravitational-wave (GW) astrophysics is successfully gearing up and is producing a wealth of landmark results, such as 
the gravitational-wave transient catalog (GWTC-1) of compact binary coalescences (CBC)  
 \cite{GWTC-1} and the tests of gravity in previously unaccessible regimes \cite{TGR}, from the recent observing runs of the Advanced LIGO \cite{LIGO} and Advanced Virgo \cite{Virgo} detectors. These important achievements come in large part from Bayesian inference applied to template-based methods \cite{Biwer_2019}. In particular, these analyses are unveiling the intrinsic properties of the binary black hole mergers, such as masses and spins, thanks to their well-understood theoretical models and their clearly recognizable waveforms (see, e.g., \cite{Cutler:1993aa,Khan:2016aa,Khan:2018fmp}).

In our work, we investigate the morphology of the events from the GWTC-1 catalog of compact binary coalescences as reconstructed by an alternative approach that does not depend on specific models, but rather extracts the coherent response of the detector network to generic gravitational waves and uses very loose additional bounds on the duration and bandwidth of the signal. 
This methodology is robust with respect to a variety of well-known CBC signal features including the higher order modes \cite{Calderon:2018}, high mass ratios, misaligned spins and eccentric orbits: it complements the existing template-based algorithms by searching for new and possibly unexpected CBC populations and waveform features. 
The present analysis extends the well-established analyses reported in \cite{TGR, GWTC-1} by looking for features that add to the conventional CBC signal models, especially at times after coalescence.

In section \ref{cWB}, we briefly introduce \textit{coherent WaveBurst} (cWB), an unmodeled algorithm for searching and reconstructing GW transients and its open source version \footnote{cWB home page, \url{https://gwburst.gitlab.io/}; \\ public repositories, \url{https://gitlab.com/gwburst/public}\\ documentation, \url{https://gwburst.gitlab.io/documentation/latest/html/index.html}.} which is adopted for the production of all results reported in this paper. 
Section \ref{MTC} describes the Monte Carlo simulation 
which enables a quantitative evaluation of the consistency between the waveform estimates by cWB versus posterior samples from template-based Bayesian inference.  
Section \ref{results} illustrates the actual cWB results for the set of GW events from 
the GWTC-1 \cite{GWTC-1}\footnote{The LIGO-Virgo data are publicly available at the Gravitational Wave Open Science Center \cite{GWOSC}; the LIGO-Virgo release of waveform posterior samples is at  \url{https://dcc.ligo.org/LIGO-P1800370/public}.}
and provides the comparison of reconstructed waveforms and sky localizations to those provided by methods based on signal templates. Follow-up studies on the two main outliers (the events GW151012 and GW151226), where cWB detects an excess of coherent energy after coalescence, are reported in section \ref{gw151012}.  
Finally, in section \ref{discussion} we discuss plausible interpretations for the observed deviations in the reconstructed waveforms for GW151012 and for GW151226. We conclude the paper with a very brief discussion 
of the perspectives for further studies that are enabled by our methods and of the opportunities of their implementation in the ongoing LIGO-Virgo survey \cite{ObservingScenario}.


\section{Methods\label{methods}}
\subsection{coherent WaveBurst \label{cWB}}
Coherent Waveburst (cWB)~\cite{KLIMENKO} is an analysis  pipeline used in searches for generic transient signals with networks of GW detectors. Designed to operate without a specific waveform model, cWB first identifies coincident excess power in the multi-resolution time-frequency representations of the detectors strain data. It then reconstructs events which are coherent in multiple detectors as well as  the source sky location and signal waveform: by using the constrained maximum likelihood method \cite{Klimenko:2008fu}, it combines all data streams into one coherent statistic $\eta_c$, which is then used for ranking cWB events. This statistic is based on the coherent energy $E_C$, such that $\eta_c \propto \sqrt{E_C}$, and it is proportional to the coherent signal-to-noise ratio (SNR) in the detector network. 
To be robust against the non-stationary detector noise, cWB employs signal independent vetoes, which reduce the initial high rate of the  excess power triggers. The cWB primary veto cut is on the network correlation coefficient $\mathrm{cc} = E_C/(E_C+E_N)$, where $E_N$ is the residual noise energy, estimated once the reconstructed signal is subtracted from the data \cite{Klimenko:2008fu}. Typically, for a GW signal $\mathrm{cc} \approx 1$ and for instrumental glitches $\mathrm{cc} \ll 1$. Therefore, during the last observing runs by the advanced GW detectors, events with $\mathrm{cc}< 0.7$ were rejected as potential glitches. 
 Finally, cWB  extends its capabilities by using linear combinations of wavelets basis functions providing an overcomplete representation of the signal. 
 In order to improve the detection efficiency for stellar mass BBH sources, cWB favors events with patterns where the frequency increases with time, i.e.
events with a chirping structure, which captures the phenomenological behavior of
most CBC sources. 

All results reported in this paper have been produced by the open source version 6.3.0 of cWB:
this version has been publicly released for open use and it reproduces cWB scientific results included in the most recent publications of the LIGO-Virgo collaboration based on the observing runs 1 and 2 \cite{GW170104, GW170608, GW170814, GWTC-1}. The pipeline and production parameters are the same as in the abovementioned publications with
minimal changes in the code to enable the measurement of quantities needed for the Monte Carlo.
While cWB can work with arbitrary detector networks, the waveform reconstruction has been restricted to the Hanford and Livingston detectors,
due to the limited contribution by Virgo data to the morphological reconstruction of the events; Virgo data was  included instead whenever possible in the sky localization reported in Table \ref{tab:skymaps}.

\subsection{Monte Carlo simulation methodology \label{MTC}}
 The Monte Carlo simulation described in this Section provides the quantitative evaluation of the consistency between the waveform estimates by cWB versus the distribution of waveform posterior samples from template-based Bayesian inference. This method is an evolution of what already used to estimate the uncertainty on the reconstructed waveforms in section IV of the GWTC-1 catalog paper \cite{GWTC-1}.
 We use the same public parameter estimation samples by LALInference \cite{Veitch:2014wba}, released for GWTC-1. For each GW event, we repeatedly add to the actual data stream a waveform extracted at random from the parameter estimation posteriors and perform the signal injection in an off-source period close to the selected event. Next, for each such injection we perform an unmodeled reconstruction and determine the maximum likelihood point-like estimated waveform and the full likelihood sky map. The final result of the Monte Carlo is an empirical distribution of cWB waveform reconstructions, which is marginalized with respect to the selected parameter estimation posteriors (in this case, with respect to LALInference parameter estimations) and which includes all effects due to non-Gaussian noise fluctuations as well as all biases due to cWB estimation itself.  These distributions of waveform and sky localization reconstructions allow the estimation of frequentist confidence intervals and the likelihood sky map coverage. For more details on the actual implementation of the Monte Carlo, see Appendix \ref{A:MTC}.

\section{Results \label{results}}

The Advanced LIGO \cite{LIGO} detectors started taking data on September $12^{th}$, 2015 until January $19^{th}$, 2016. This run is referred to as O1.
The Advanced LIGO detectors started their second run, called O2, on November $30^{th}$, 2016, and on the first of August, 2017 the Advanced Virgo \cite{Virgo} detector joined the
observing run, enabling the first three-detector observations of
GWs. This second run ended on August $25^{th}$, 2017.
During the first and the second runs,  
the LIGO-Virgo searches for binary mergers (both those with matched filters and unmodeled) identified a total of ten BBH mergers and
one binary neutron star (BNS) signal. 
All relevant information on those GW events can be found in the Gravitational-Wave Transient Catalog of Compact Binary Mergers (GWTC$-$1) \cite{GWTC-1} and in a number of other papers dedicated to individual GW events \cite{GW150914, GW151226, GW170104, GW170608, GW170814, GW170817}. Moreover, information on O1 GW events, including a large set of subthreshold candidates, can be found in \cite{Nitz_2019}. Finally, previous tests checking for deviations from CBC models have been discussed in \cite{TGR}. 

\begin{figure*}[h]
\subfloat[]{\includegraphics[clip,width=0.33\textwidth]{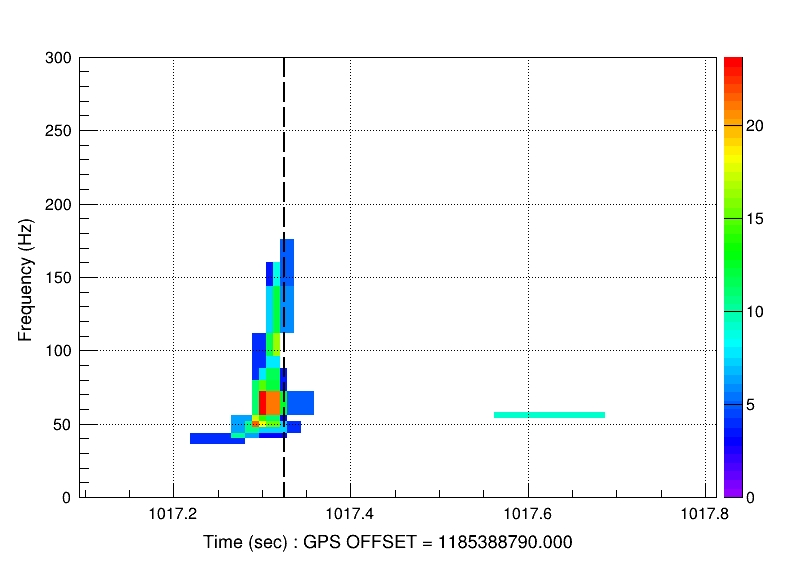}}\hfil
\subfloat[]{\includegraphics[clip,width=0.33\textwidth]{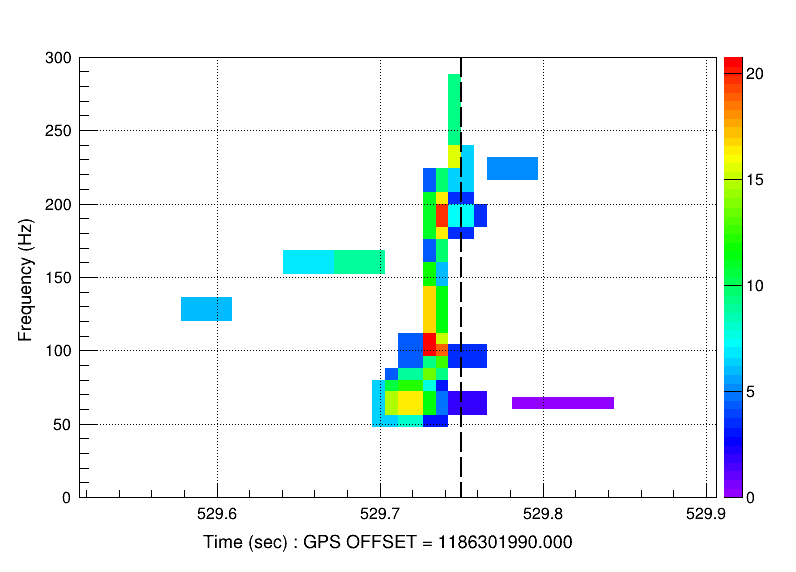}}
\hfil
\subfloat[]{\includegraphics[clip,width=0.33\textwidth]{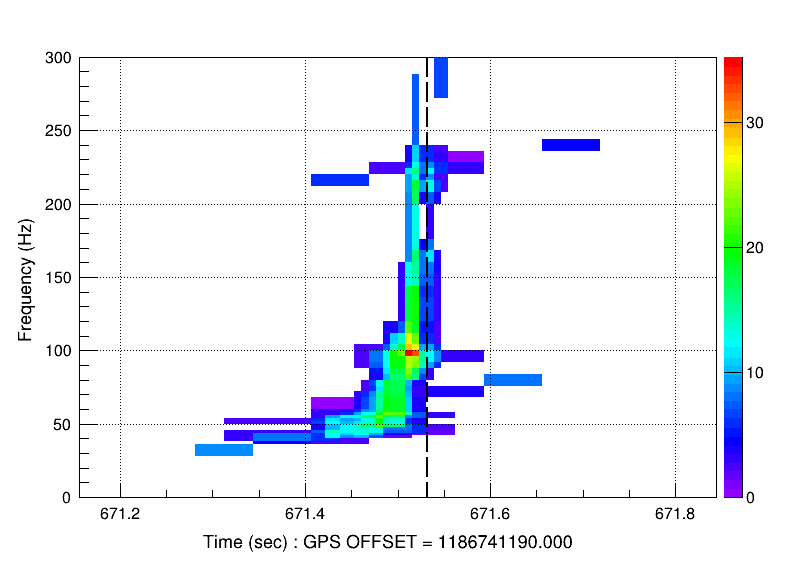}}

\subfloat[]{\includegraphics[clip,width=0.33\textwidth]{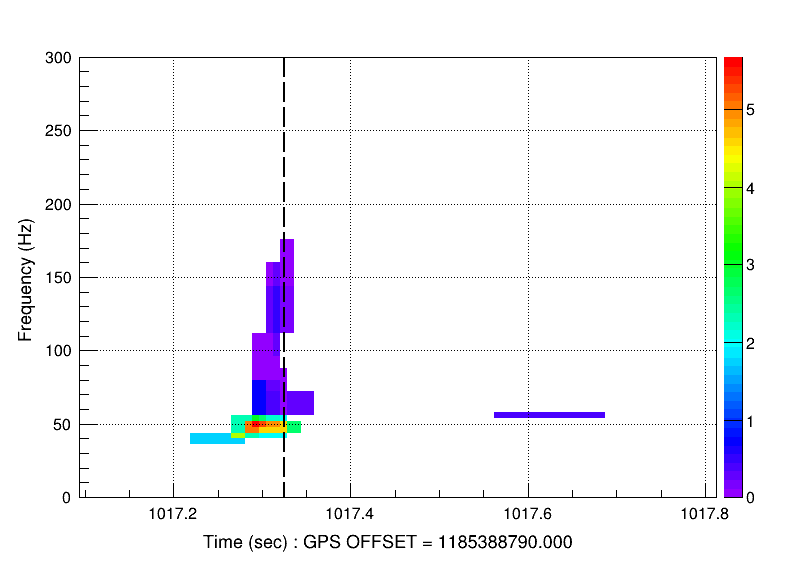}}\hfil
\subfloat[]{\includegraphics[clip,width=0.33\textwidth]{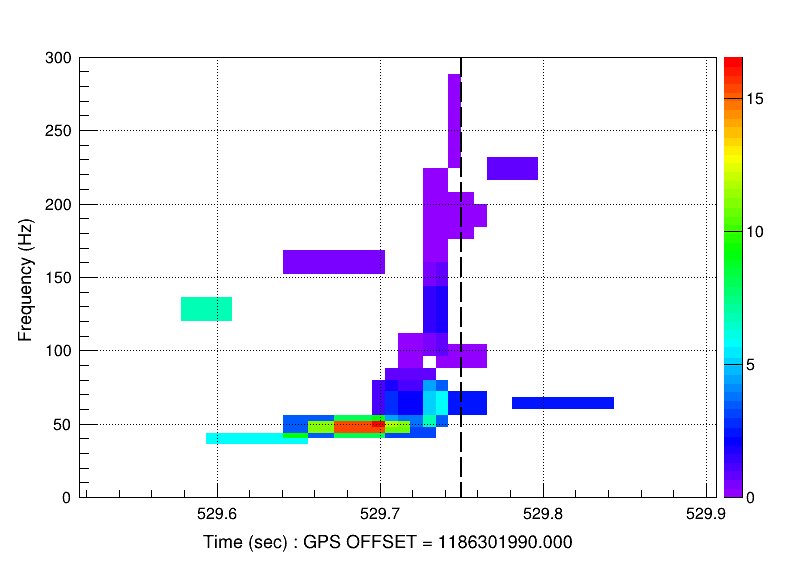}}
\hfil
\subfloat[]{\includegraphics[clip,width=0.33\textwidth]{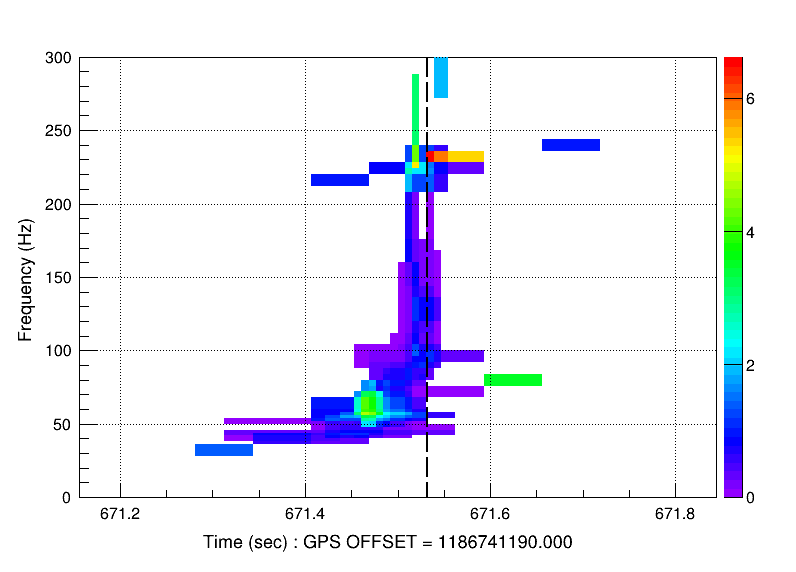}}

\caption{\label{f:TFexamples} cWB waveform reconstruction of GW170729, GW170809 and GW170814 in form of color-coded time-frequency maps. The upper row shows the squared coherent network SNR, while the lower row shows the  normalized residual noise energy, $E_{N}$, estimated after the reconstructed signal is subtracted from the data. The first column ((a) and (d)) refers to GW170729; next, the  second column ((b) and (e)) shows event GW170809; finally, the third column shows the event GW170814. The dashed vertical lines denote the minR $\mathrm{t_{L,coa}}$ for these three events (the network reconstruction uses the Livingston detector time as a reference).}
\end{figure*}

When considering cWB reconstruction of the eleven main events from the GWTC$-$1, a subset (i.e. GW151012, GW151226, GW170729, GW170809 and GW170814) displays features in well-localized regions of the time-frequency representation, mainly close to the merger time, which are missing in the usual CBC waveforms (see Figure \ref{f:TFexamples}, Figures \ref{f:TF1} (a) and (d) and Figures \ref{f:TF2} (a) and (d)).

\subsection{cWB unmodeled reconstruction of the events within GWTC--1 \label{GWTC-1}}

For each GW event, we compared the cWB sky localization likelihood map with the corresponding posterior probability sky regions produced by LALInference \cite{Veitch:2014wba}: while the unmodeled reconstruction produces more extended and irregular sky regions and different biases may arise \footnote{cWB customarily employs an antenna pattern prior, which overweights favorable incoming directions, while underweighting the ones with poor
antenna pattern. This affects especially two-detector networks, as it the case for this work.}, the overall overlap between the estimates by cWB and  
LALInference is mostly good. More details can be found in appendix \ref{A:skymaps}, in particular in Table \ref{tab:skymaps} and in Figure \ref{Mollweide}.

Similarly, for each GW event, the waveforms reconstructed by cWB are compared with those reconstructed from the Monte Carlo which uses LALInference posterior samples: as expected from an algorithm based on excess power, the merger part of the signal is usually well estimated by cWB, while the early inspiral and the ringdown parts (where the signals are weaker and spread over larger areas in the time-frequency plane) are often missing.

We define the corresponding CBC whitened waveforms, 
$\mathbf{w}$ = $[w_{H}(t),w_{L}(t)]$,
starting from the CBC modeled strain waveforms at the Hanford and Livingston sites, $\mathbf{h}$ = $[h_{H}(t),h_{L}(t)]$ and dividing them in the frequency domain by the cWB-estimated noise spectral densities, $\sqrt{{S_{n,(H,L)}(f)}}$. By denoting the model-independent whitened cWB waveforms as $\hat{\mathbf{w}}$, we can calculate the generalized fitting factor of $\hat{\mathbf{w}}$ with respect to $\mathbf{w}$ as  

\begin{equation}
 \mathrm{FF}(\mathbf{w},\hat{\mathbf{w}}) =  \frac{(\mathbf{w}|\hat{\mathbf{w}})}{\sqrt{(\mathbf{w}|\mathbf{w})}\sqrt{(\hat{\mathbf{w}}|\hat{\mathbf{w}})}}
\end{equation}

where the scalar product is defined in the time domain as  

\begin{equation}
 (\mathbf{w}|\hat{\mathbf{w}}) = \int_{\mathbf{t_{1}}}^{\mathbf{t_{coa}}} \mathbf{w}(t) \hat\mathbf{w}(t) dt
\end{equation}

with $\mathbf{t}_{1}=[t_{H,1},t_{L,1}]$ defined as $\int_{-\infty}^{\mathbf{t}_1} \left|\hat\mathbf{w}(t)
\right|^2 dt = 0.01 \int_{-\infty}^{\infty} \left|\hat\mathbf{w}(t)
\right|^2 dt $ and $\mathbf{t_{coa}}=[\mathrm{t_{H, coa}},\mathrm{t_{L,coa}}]$ 
is the coalescence time \footnote{In this article, we considered just the IMRPhenomP waveform family, where the $\mathrm{t_{coa}}$ is the estimated time (with a stationary phase approximation)  corresponding to the $\mathrm{f_{peak}}$ as defined in Eq. 20 in \cite{Khan:2016aa}. }.

In Table \ref{tab1}, we report the average $\langle \mathrm{FF}\rangle_\mathrm{offsource}$ calculated by Monte Carlo sampling, where for each injected LALInference posterior, $\mathbf{h}{^{i}(t)}$ (the index $i$ runs over all posterior samples; see Appendix \ref{MTC} for more details) we calculated $\mathrm{FF}(\mathbf{w}^{i}(t),\hat{\mathbf{w}}^{i}(t))$, as well as the ``onsource'' average FF, i.e. $\langle \mathrm{FF}\rangle_\mathrm{onsource} = \langle \mathrm{FF}(\mathbf{w}^{i}{(t)},\hat{\mathbf{w}}^\mathrm{gw}(t))\rangle_{i}$, where $\hat{\mathbf{w}}^\mathrm{gw}(t)$ is the actual cWB whitened waveform of the GW event. 

 By defining the residual energy of $\hat{\mathbf{w}}^\mathrm{gw}$ with respect to $i^{th}$ CBC whitened sample $\mathbf{w}^{i}$ as 
 
 \begin{equation}
 \mathbf{E}^{i}_\mathrm{res} = \int_{\mathbf{t}^i_A}^{\mathbf{t}^i_B} \left|\mathbf{w}^i(t)-\hat{\mathbf{w}}^\mathrm{gw}(t)\right|^2 dt 
 \end{equation}

  with $\mathbf{t}_{A,B}$ defined as $\int_{-\infty}^{\mathbf{t}_A} \left|\hat{\mathbf{w}}^{gw}(t)
\right|^2 dt=\int_{\mathbf{t}_B}^{\infty} \left|\hat{\mathbf{w}}^{gw}(t)
\right|^2 dt=0.02 \int_{-\infty}^{\infty} \left|\hat{\mathbf{w}}^{gw}(t)
\right|^2 dt $, we can then calculate the sample $i$ corresponding to  $\mathrm{\min}_i\left (\left|\left| \mathbf{E}^{i}_\mathrm{res} \right |\right| \right )$ and denote it as the minimal residual energy posterior sample, minR.

The relatively high $\langle \mathrm{FF}\rangle_\mathrm{offsource}$ suggests that cWB is good at reconstructing the injected waveforms before coalescence time. 
Moreover, from the comparison of the onsource and offsource FFs, we note there are no significant differences between the two cases and we infer a qualitative compatibility between estimates, with the  exception of GW170809, which we shall investigate further in the future.     
 
\subsection{Significance of post-coalescence excesses \label{pcmethod}}
A number of cWB excesses with respect to the CBC templates occur after the coalescence, where CBC waveforms are mostly silent. In order to estimate their significance, we devised a procedure based on their coherent network SNR. We considered the coalescence time, i.e. $t_\mathrm{coa}$, as a reference time to start integrating our excesses: we then split our reconstructed waveforms in a \textit{primary} part (i.e. for $t\leq t_\mathrm{coa}$) and a \textit{secondary} part (i.e. for $t> t_\mathrm{coa}$). Our main test-statistic for the secondary is then the signal-to-noise ratio
\begin{equation}
 \mathrm{SNR}_\mathrm{pc} = \sqrt{\int_{\mathbf{t}_{\mathrm{coa}}}^{\infty} \left|\hat{\mathbf{w}}(t)\right|^2 dt}
 \end{equation}

\begin{figure*}
\includegraphics[width=0.75\textwidth]{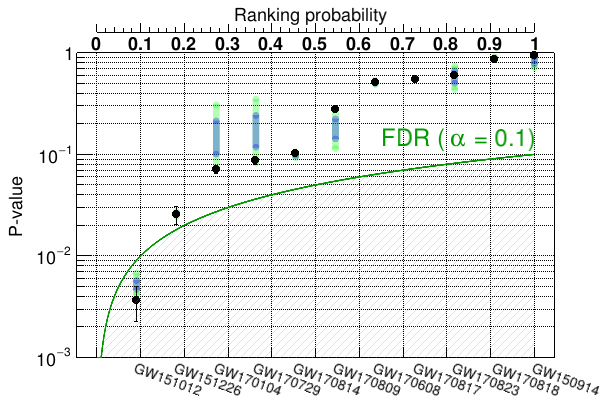}
\caption{\label{f:FDR} Sorted p-values of the post-coalescence excesses as a function of the ranking probability ($\mathrm{rank}/N_\mathrm{events}$), compared with an FDR ($\alpha=0.1$). Out of the 11 GW events, the post-coalescence $\mathrm{p}$-$\mathrm{value}_\mathrm{pc}$ for GW151012 is an outlier. The second ranked event, GW151226, though mildly significant is consistent with the 11 trials.     }
\end{figure*}

\footnote{With the previous definition of the $\mathbf{w}$s, which includes a division by the amplitude spectral density in the frequency domain, this is indeed an SNR. We also note that the resulting $\mathrm{SNR}_\mathrm{pc}$ also contains  the ringdown of the primary signal.}. While for the offsource Monte Carlo simulations, the $t_\mathrm{coa}$ are known from the CBC posterior samples, for the onsource the $t_\mathrm{coa}$ can only be estimated, leaving a margin for uncertainty on the onsource $\mathrm{SNR}_{pc}$. We decided to adopt $t^\mathrm{minR}_\mathrm{coa}$, i.e. the coalescence time of the minimal residual energy CBC posterior sample (already defined in Section \ref{GWTC-1}) as our best point estimate. On Table\ref{tab1}, we report the $\mathrm{SNR}^\mathrm{minR}_\mathrm{pc}$, together with the upper and lower bounds, $\mathrm{SNR}^\mathrm{sup}_\mathrm{pc}$ and $\mathrm{SNR}^\mathrm{inf}_\mathrm{pc}$ which correspond to $\langle\mathrm{t}_\mathrm{coa}\rangle \pm2\sigma$. Finally, in the last column of Table \ref{tab1}, we report the estimated p-value for the onsource coherent SNR excess as the ratio of the number of offsource reconstructed waveforms with a larger or equal value of coherent $\mathrm{SNR}_\mathrm{pc}$ over the total number of Monte Carlo reconstructed injections. Consistently with the definition of $2 \sigma $ SNR upper and lower bounds, we report the upper and lower bounds for the p-value.
In order to evaluate the result in the context of multiple hypothesis tests, i.e. when comparing many data against a given hypothesis, we adopt the false-discovery rate (FDR) procedure to control the number of mistakes made
when selecting potential deviations from the null hypothesis \cite{Miller_2001}:
 the result is shown in Figure \ref{f:FDR} where the sorted post-coalescence p-values (black dots represent the minR p-values, while the blue and light green vertical lines show the 1 and 2$\sigma$ [$P_\mathrm{sup}$-$P_\mathrm{inf}$] bounds, respectively) are plotted against the ranking probability ($\mathrm{rank}/N_\mathrm{events}$) and compared with an FDR light green dashed area (with false alarm probability $\alpha=0.1$). Out of the 11 GW events, the post-coalescence $\mathrm{p}$-$\mathrm{value}_\mathrm{pc}$ for GW151012 is an outlier. The second ranked event, GW151226, though mildly significant is consistent with the 11 trials.

\begin{table*}[htbp]
\label{table1}
 \centering
 \begin{tabular}{l c c c c c c c} 
  \toprule
    {\specialcellbold{GW event }}                 &
    {\specialcellbold{source \\ }}                 &
    {\specialcellbold{$\langle \mathrm{FF}\rangle_\mathrm{offsource}$ \\ }}                  &
     {\specialcellbold{$\langle \mathrm{FF}\rangle_\mathrm{onsource}$ \\ }}                  &
    {\specialcellbold{$\mathrm{FF}^\mathrm{minR}$ \\ }}                 &
    {\specialcellbold{$\mathrm{SNR}$}}                 &
    {\specialcellbold{$\mathrm{SNR}^\mathrm{minR}_\mathrm{pc} \left \{ ^{\mathrm{SNR}^\mathrm{sup}_\mathrm{pc}}_{\mathrm{SNR}^\mathrm{inf}_\mathrm{pc}} \right \} $} }         &
    {\specialcellbold{{$\mathrm{p}$-$\mathrm{value}$}$_\mathrm{pc} \left \{^{P_\mathrm{sup}}_{P_\mathrm{inf}}\right \}$\\ } }      \\
   \midrule
  GW150914 & BBH & $0.95\pm0.02$ & $0.96\pm0.01$ & $0.97$ & $25.2$ &  $5.72~^{6.92}_{5.64}$ & $0.94\pm0.02~^{0.95}_{0.71}$ \\
 GW151012 & BBH &  $0.80\pm0.10$  & $0.82\pm0.038$ & $0.9$ & $ 10.5$ &  $6.60~^{6.54}_{6.26}$ & $0.0037\pm0.0014~^{0.0068}_{0.0042}$ \\
GW151226 & BBH & $0.78\pm0.08$ & $0.75\pm0.05$ & $0.85$ & $ 11.9$ &  $4.40~^{4.41}_{4.36}$ & $0.025\pm0.005~^{0.03}_{0.02}$ \\
GW170104 & BBH & $0.90\pm0.05$ & $0.95\pm0.03$ & $0.97$ & $ 13.0$ &  $5.29~^{5.30}_{3.95}$ & $0.07\pm0.01~^{0.31}_{0.07}$ \\
GW170608 & BBH & $0.79\pm0.07$ & $0.81\pm0.02$ & $0.84$ & $ 14.1$ &  $1.69~^{1.75}_{1.64}$ & $0.51\pm0.02~^{0.54}_{0.49}$ \\
GW170729 & BBH & $0.90\pm0.05$ & $0.93\pm0.02$ & $0.95$ & $ 10.2$ &  $4.81~^{4.86}_{3.43}$ & $0.09\pm0.01~^{0.35}_{0.08}$ \\
GW170809 & BBH & $0.90\pm0.04$ & $0.78\pm0.03$ & $0.82$ & $ 11.9$ &  $3.89~^{4.71}_{3.88}$ & $0.28\pm0.01~^{0.28}_{0.11}$ \\
GW170814 & BBH & $0.92\pm0.03$ & $0.91\pm0.02$ & $0.93$ & $ 17.2$ &  $5.98~^{6.02}_{5.94}$ & $0.10\pm0.01~^{0.11}_{0.09}$ \\
GW170817 & BNS & $0.78\pm0.05$ & $0.7596\pm0.0004$ & $0.76$ & $ 29.3$ &  $0.21~^{0.21}_{0.21}$ & $0.55\pm0.01~^{0.56}_{0.55}$ \\
GW170818 & BBH & $0.89\pm0.06$ & $0.87\pm0.01$ & $0.92$ & $ 8.6$ &  $1.97~^{2.04}_{1.76}$ & $0.87\pm0.02~^{0.91}_{0.86}$ \\
GW170823 & BBH & $0.91\pm0.05$ & $0.96\pm0.03$ & $0.98$ & $ 10.8$ &  $3.11~^{3.54}_{2.69}$ & $0.60\pm0.02~^{0.74}_{0.44}$ \\
\bottomrule
\end{tabular}
\caption{\label{tab1}Significance of post-coalescence deviations recostructed by cWB on the eleven GW events from GWTC-1. For each event we report: offsource and onsource average fitting factors, FF, with 1$\sigma$ uncertainties; the onsource FF corresponding to the minimal residual energy posterior sample; the network SNR as estimated by cWB; the post-coalescence SNR, $\mathrm{SNR}^\mathrm{minR}_\mathrm{pc}$, assuming the $t_\mathrm{coa}$ from the minR posterior sample (while $\mathrm{SNR}^\mathrm{sup}_\mathrm{pc}$ and $\mathrm{SNR}^\mathrm{inf}_\mathrm{pc}$ are the onsource $\mathrm{SNR}_\mathrm{pc}$ corresponding to $\langle\mathrm{t}_\mathrm{coa}\rangle$ $\pm 2\sigma$, respectively); and finally, the empirically estimated probability that such an $\mathrm{SNR}^\mathrm{minR}_\mathrm{pc}$ may be produced by a noise fluctuation (where the $P_\mathrm{sup}$ and $P_\mathrm{inf}$ values refer to the probability of $\mathrm{SNR}^\mathrm{inf}_\mathrm{pc}$ and $\mathrm{SNR}^\mathrm{sup}_\mathrm{pc}$, respectively).}
\end{table*}

\subsection{Follow-ups on GW151012 and GW151226 }
\label{gw151012}
\begin{figure*}[h]
\subfloat[]{\includegraphics[clip,width=0.33\textwidth]{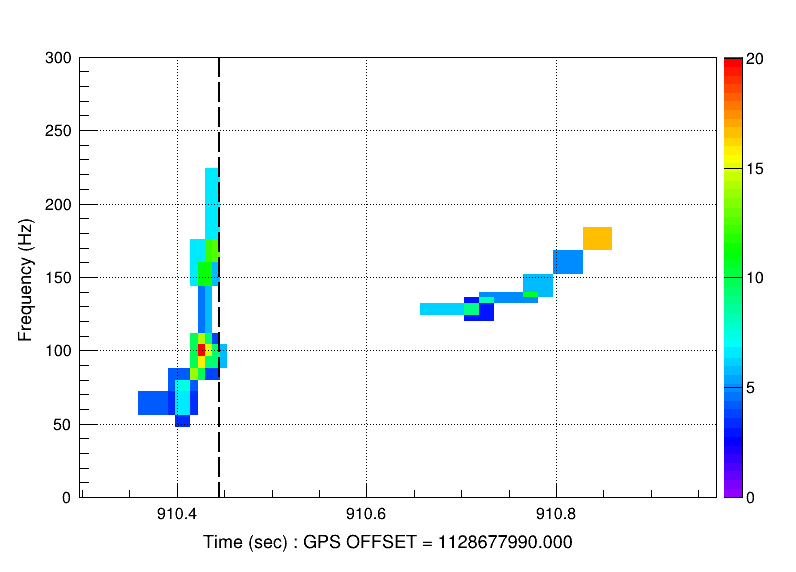}}\hfil
\subfloat[]{\includegraphics[clip,width=0.33\textwidth]{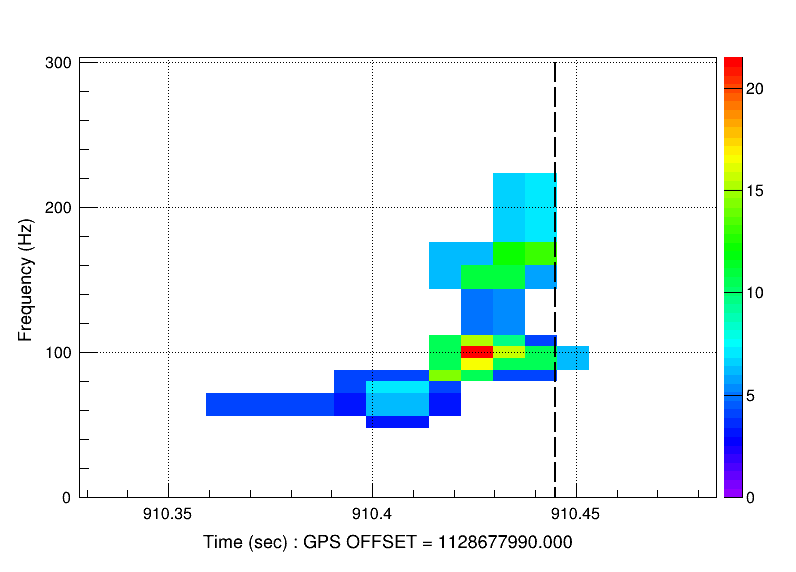}}
\hfil
\subfloat[]{\includegraphics[clip,width=0.33\textwidth]{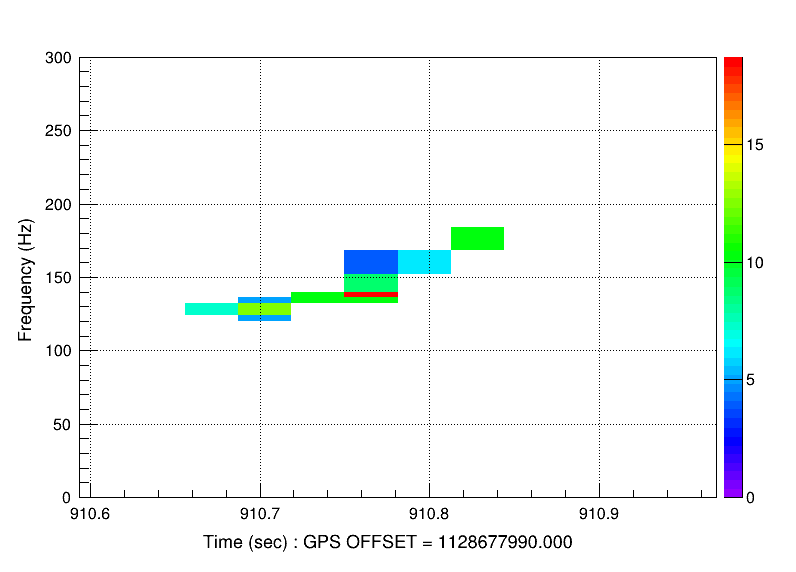}}

\subfloat[]{\includegraphics[clip,width=0.33\textwidth]{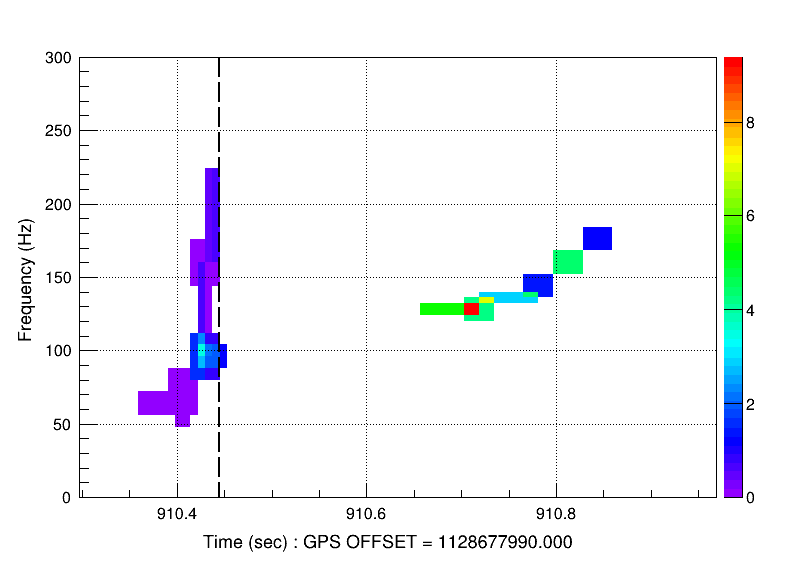}}\hfil
\subfloat[]{\includegraphics[clip,width=0.33\textwidth]{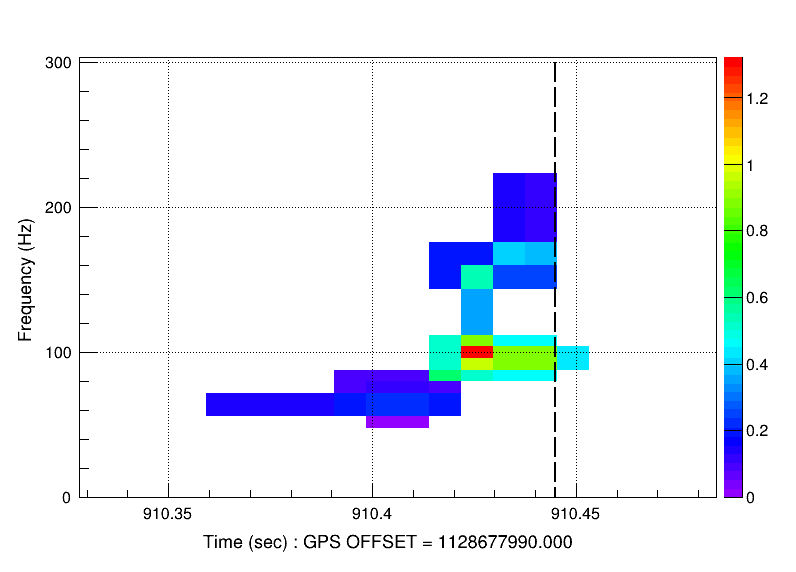}}
\hfil
\subfloat[]{\includegraphics[clip,width=0.33\textwidth]{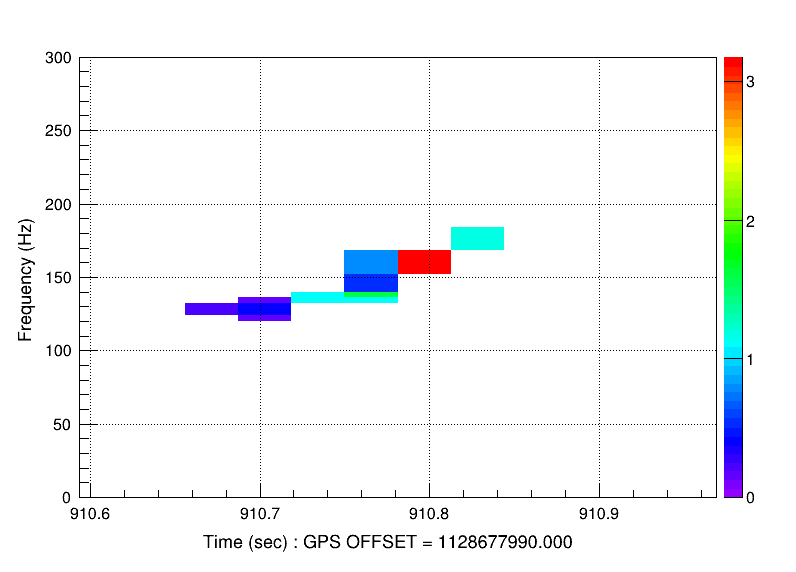}}

\caption{\label{f:TF1} cWB waveform reconstruction of GW151012, in form of color-coded time-frequency maps. The upper row shows the squared coherent network SNR,  while the second row shows the normalized residual noise energy, $E_{N}$, estimated after the reconstructed signal is subtracted from the data. The first column ((a) and (d)) refers to the original reconstruction, with the primary chirp on the left matching the CBC PE reconstruction and a secondary cluster occurring $~200$ ms after the merger; an \textit{ad hoc} time veto covering the secondary cluster was used to produce our best estimate for GW151012 primary event shown in the second column ((b) and (e)); finally, the third column reports the independent reconstruction of the secondary cluster by vetoing the primary event. 
The dashed vertical lines denote the minR $\mathrm{t_{L,coa}}$ for GW151012 (the network reconstruction uses the Livingston detector time as a reference)}
\end{figure*}

\begin{figure*}[h]
\subfloat[]{
\includegraphics[clip,width=0.45\textwidth]{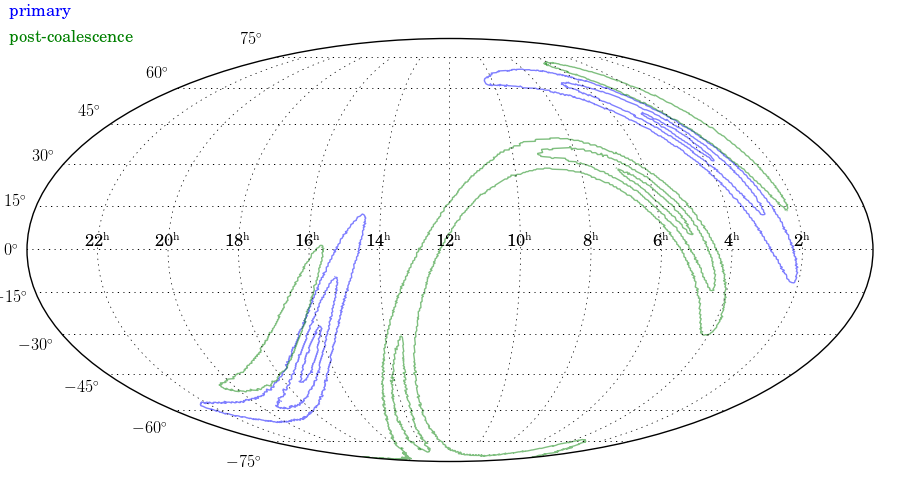}
}
\hfil
\subfloat[]{
\includegraphics[clip,width=0.45\textwidth]{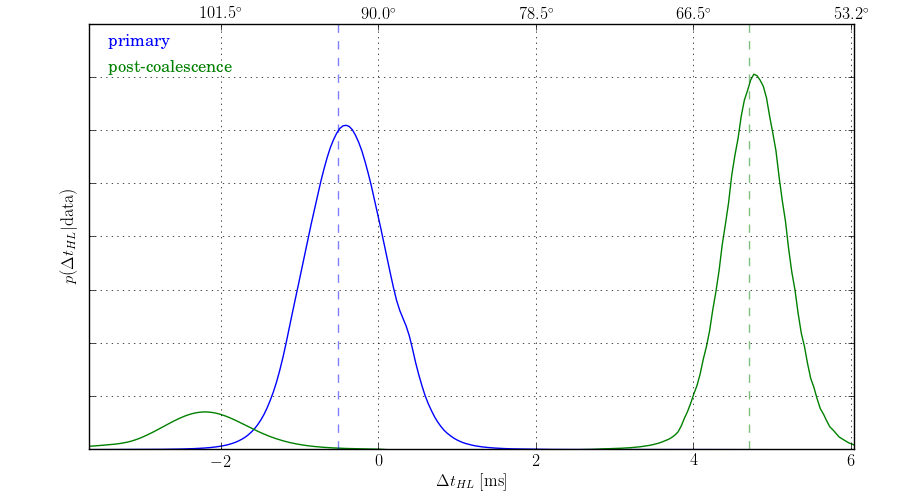}
}

\caption{\label{f:sky1} Left panel: Mollweide projection of GW151012 primary event (blue contour lines) and the secondary event (green lines) sky reconstructions in equatorial coordinates (a). Right panel: 
time delay maximum \textit{a posteriori} probability marginals between H and L in line-of-sight frame defined by H and L (b). Both figures have been produced using the code in \cite{reedessick}. }
\end{figure*}

The original reconstruction of GW151012 reveals a primary part consistent with the time-frequency evolution estimated by LALInference, followed by a secondary part, with a chirping structure, at roughly 200 ms after the merger of GW151012, as shown in Figure \ref{f:TF1} (a). The $E_{N}$ in Figure \ref{f:TF1} (d) shows that the secondary part, when reconstructed together with the primary, has some relatively high residuals.  
 In order to produce the plots in the second and third columns of Figure \ref{f:TF1}, we applied \textit{ad hoc} time vetoes to reconstruct indipendently the primary ((b) and (e)) and secondary ((c) and (f)). As we see from Figure \ref{f:sky1} (a), the estimated sky localizations differ substantially between the primary (blue) and the secondary (green),  when reconstructed independently; the main reason being the reconstructed time delay between the detectors (Figure \ref{f:sky1} (b)). This suggests that the primary and secondary might be unrelated.
 
\begin{figure*}[h]
\subfloat[]{\includegraphics[clip,width=0.33\textwidth]{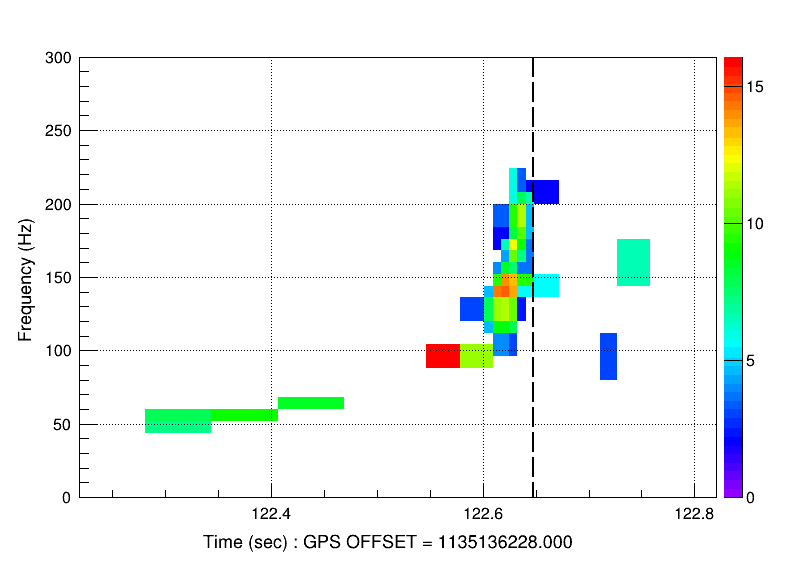}}\hfil
\subfloat[]{\includegraphics[clip,width=0.33\textwidth]{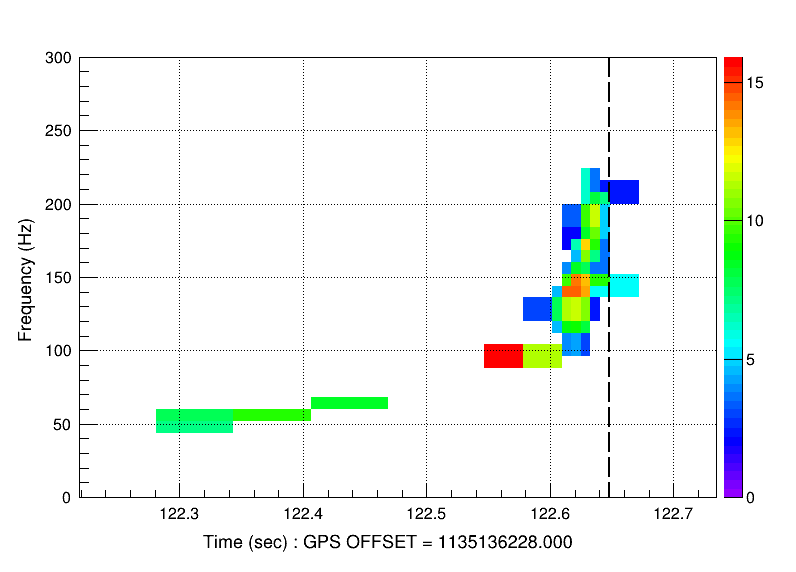}}
\hfil
\subfloat[]{\includegraphics[clip,width=0.33\textwidth]{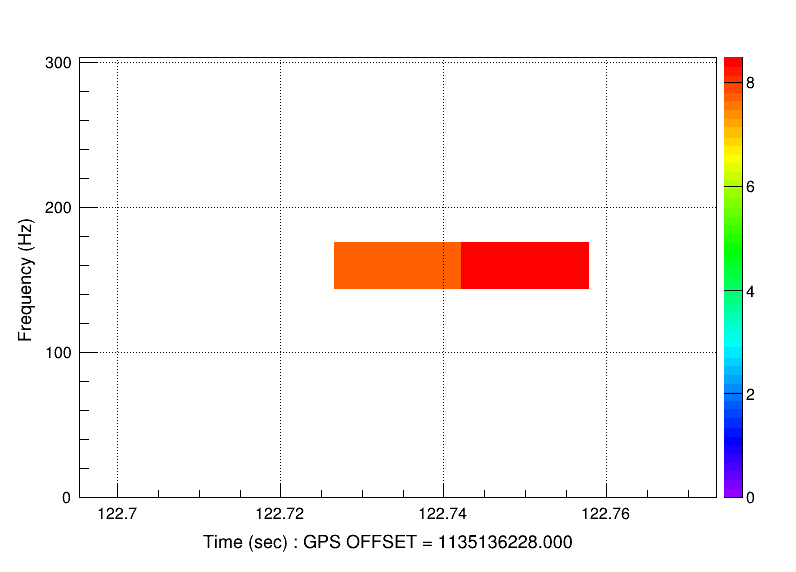}}

\subfloat[]{\includegraphics[clip,width=0.33\textwidth]{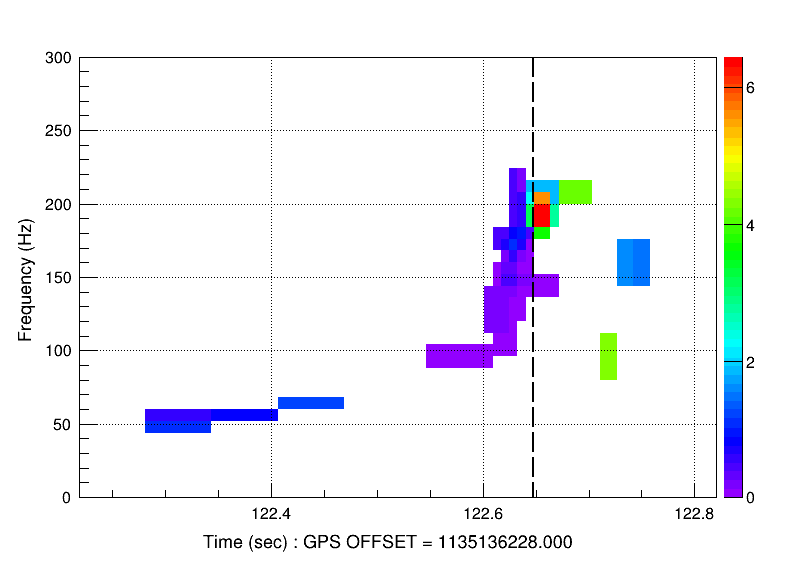}}\hfil
\subfloat[]{\includegraphics[clip,width=0.33\textwidth]{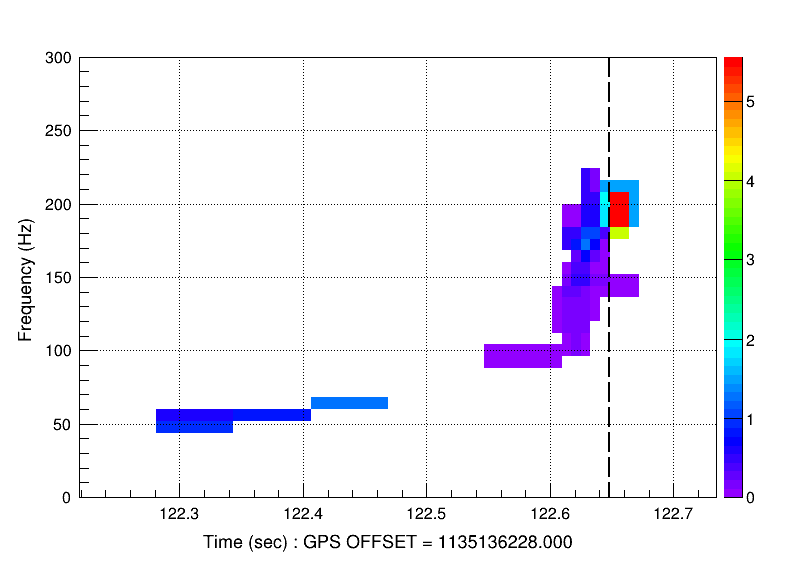}}
\hfil
\subfloat[]{\includegraphics[clip,width=0.33\textwidth]{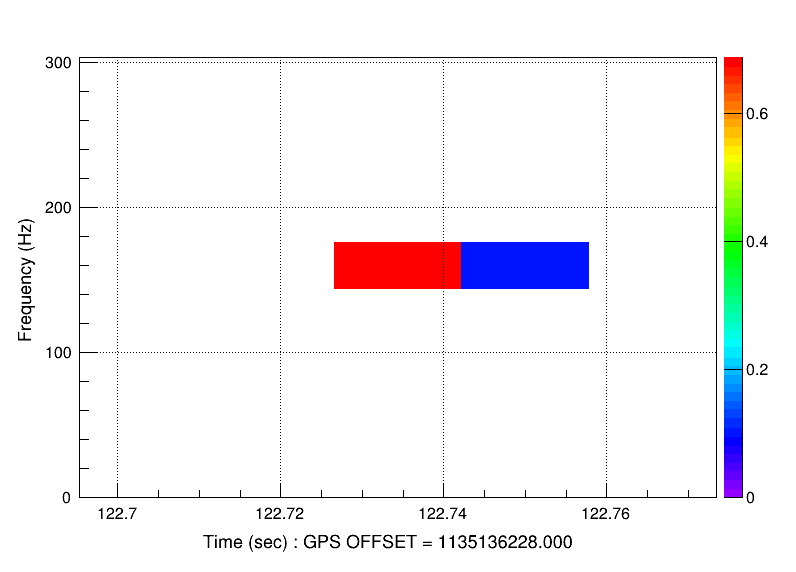}}

\caption{\label{f:TF2} cWB waveform reconstruction of GW151226, in form of color-coded time-frequency maps. The upper row shows the square coherent network SNR, while the second row shows the normalized residual noise energy, $E_{N}$, estimated after the reconstructed signal is subtracted from the data. The first column ((a) and (d)) refers to the original reconstruction, with the primary chirp on the left matching the CBC PE reconstruction and a secondary cluster occurring $~100$~ms after the merger. An \textit{ad hoc} time veto covering the secondary cluster was used to produce our best estimate for GW151226 primary event shown in the second column ((b) and (e)). Finally, the third column reports the independent reconstruction of the secondary cluster by vetoing the primary event. The dashed vertical lines show the minR $\mathrm{t_{L,coa}}$ for GW151012 (the network reconstruction uses the Livingston detector time as a reference).}
\end{figure*}

\begin{figure*}[h]
\subfloat[]{
\includegraphics[clip,width=0.45\textwidth]{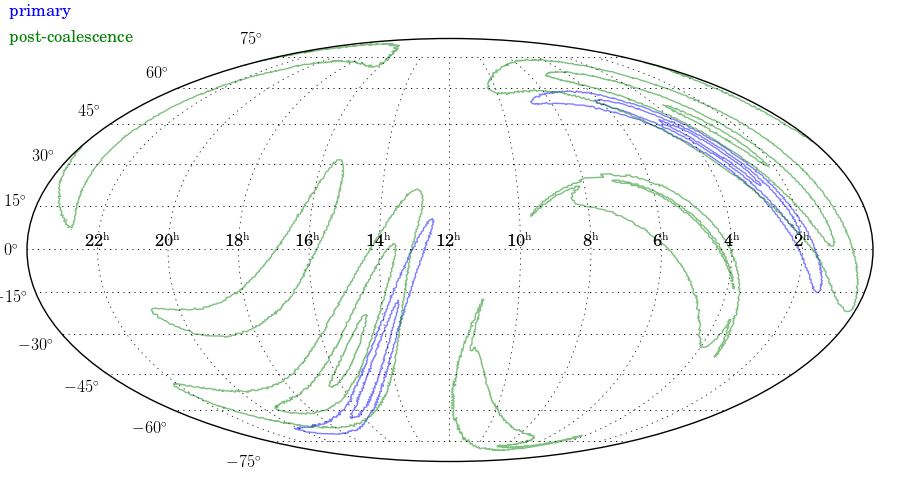}
}
\hfil
\subfloat[]{
\includegraphics[clip,width=0.45\textwidth]{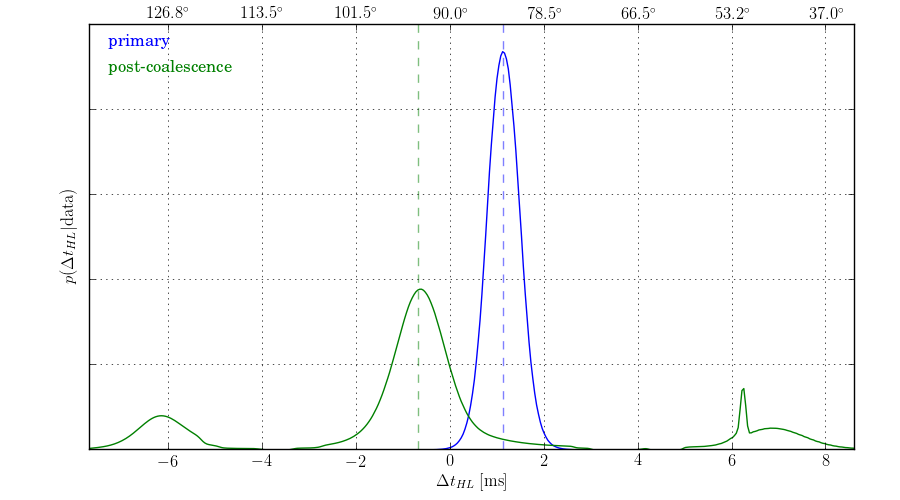}
}

\caption{\label{f:sky2} Left panel: Mollweide projection of GW151226 primary event (blue contour lines) and the secondary event (green lines) sky reconstructions in equatorial coordinates (a). Right panel: 
time delay maximum \textit{a posteriori} probability marginals between H and L in line-of-sight frame defined by H and L (b). Both figures have been produced using the code in \cite{reedessick}. }
\end{figure*}

 A similar follow-up analysis was conducted also on GW151226. In Figures \ref{f:TF2} (a) and (d) the secondary is visible about 100 ms after coalescence. The plots in the second and third columns of Figure \ref{f:TF2}, show the primary ((b) and (e)) and secondary ((c) and (f)) reconstructed indipendently of one another; the latter being very weak and lacking any structure. Figures \ref{f:sky2} (a) and (b) show that the estimated sky localizations of the primary (blue) and the secondary (green) could be compatible.

 In Table \ref{tab:follow-ups}, we report the follow-up parameters for GW151012 and GW151226 in the original reconstruction and by isolating the primary and the secondary: the network SNR, the correlation coefficient (defined in Section \ref{cWB}) and the chirp mass, as estimated by LALInference (PE) and by cWB \cite{Tiwari_2015}. When comparing the network SNRs of the primary with the estimates by LALInference, we find an SNR loss of $\approx 10\textendash15\%$; cWB chirp mass estimates are mostly used just as an indication of ``chirpness'' as its estimator relies on simplified assumptions that can lead to  relatively large biases.    

\begin{table*}[htbp]
 \centering
 \begin{tabular}{l c c c c c c c c c c c} 
  \toprule
    {\specialcellbold{GW event }}                 &
    \multicolumn{4}{c}{SNR}           &
    \multicolumn{3}{c}{cc}      &
    \multicolumn{4}{c}{Chirp Mass [\Msun]} \\ 
    \cmidrule(lr){2-5}
    \cmidrule(lr){6-8}
    \cmidrule(lr){9-12}

             &  PE & original     &  primary     &    secondary 
             &   original     &  primary      &    secondary
             &  PE & original     &  primary     &    secondary \\
             
\midrule
GW151012 & 10.0& 10.5 & 8.4& 7.1& 0.81& 0.95& 0.83& 15.2& 23.8& 23.8& 3.2\\
GW151226 & 13.1& 11.9 & 11.4 & 4.0 & 0.82& 0.85& 0.86 &8.9 &  10.4& 10.4& ---\\
\bottomrule
\end{tabular}
\caption{\label{tab:follow-ups} Follow-up parameters for GW151012 and GW151226: the network SNR, the correlation coefficient (defined in Section \ref{cWB}) and the chirp mass as estimated by LALInference (PE) and by cWB in the original reconstruction and by isolating the primary and the secondary.}
\end{table*}

\section{Discussion \label{discussion}}

In this paragraph we focus on the two outliers of the distribution of post coalescence features from Table I, GW151012 and GW151226.  The p-values are too large to exclude the null hypothesis, and yet it is useful  to consider alternative explanations to exemplify the kind of findings that could be borne out of our methodology. 


\subsection{GW151012}

A first plausible interpretation for the primary and secondary in the cWB reconstruction of GW151012 is that their occurrence may be due to an accidental coincidence of unrelated CBC events of different chirp masses and directions, compatible with catalogs of subthreshold CBC candidates, see in particular 1-OGC \cite{Nitz_2019}. 
While the primary component is a well established CBC detection, the secondary feature on its own is much fainter and 
it is likely originated by coherent noise from the detectors.
In any case, even assuming an astrophysical origin for the secondary, according to our morphological analysis the two GWs would be unrelated.

The secondary feature shows traits which are consistent with the CBC subthreshold candidates in the 1-OGC catalog \cite{Nitz_2019}, both morphologically and in terms of signal amplitude.  In fact, its morphology is consistent with a coalescence of a light stellar mass binary, within the parameter space included by the template bank of the 1-OGC search.
Also the SNR values collected at each detector by cWB are larger than 4, the threshold used for 1-OGC, and the network SNR recovered by cWB is very close to the mode of the SNR distribution of the subthreshold events in 1-OGC. 
 \footnote{The secondary feature calls for a mass ratio of the binary component significantly different from 1, from simplified matched filter analyses} 
We expect the SNR collected by cWB to be statistically comparable with the one collected by CBC matched filter searches. 
This is tested for the detections described in the GWTC-1 catalog \cite{GWTC-1}, and we expect it to hold also at the fainter amplitudes comparable to the secondary.
The secondary feature was not found in 1-OGC search for subthresholds CBC candidates probably because of the dead time set around all reconstructed events, such as the primary \cite{Nitz_2019}. 

The rate of occurrence of subthreshold CBC candidates in 1-OGC is $\sim 0.06$~Hz, estimated from the total counts of candidates divided by the net observation time after accounting for the deadtime around each selected candidate. 
cWB clusters together two signals if their time separation is less than 0.2~s, which corresponds to opening a time window of variable width after the coalescence time, depending on the actual duration of each signal component. 
In the case of GW151012, the typical time window ranges from 0.25~s for faint BBH-like secondaries to 0.37~s for faint BNS-like secondary. 
Therefore,  the probability of random occurrence after coalescence of a subthreshold 1-OGC event within the cWB  window ranges from 1.4\% to 2.1\% during the O1 run. 
Assuming similar rates for the subthreshold events in the O2 run, the order of magnitude of this probability is high enough to allow the possible occurrence of a CBC-like secondary in one out of the 11 regions after coalescence tested here. 
On the other hand, this probability is low enough not to be inconsistent with the measured p-value of GW151012, taking into account the uncertainties on the detection efficiency loss of cWB for candidates of the class of the 1-OGC candidates after the primary coalescence. 
Therefore we conclude that the secondary feature has a plausible interpretation as an object of the class of the subthreshold CBC candidates, which in turn have a population dominated by coherent noise fluctuations.

Our morphological analysis strongly disfavors any interpretation of the post-coalescence feature that is directly related to the detected GW151012,  because of the inconsistency in the delay times at the detectors, which call for a different source position. In particular, this analysis disfavors alternative interpretations as post-merger echoes (see previous analyses in \cite{abedi2017echoes} and \cite{nielsen2018parameter}), despite an apparent agreement in the delay time from merger and amplitude. In addition, the reconstructed spectral content does not fit the proposed echoes models. 

\subsection{GW151226}

The p-value of the post-coalescence feature is only marginally interesting in our analysis, and the preferred interpretation is the null hypothesis. 
Here, the alternative, though statistically disfavored, interpretation for the  post-coalescence feature --  a single tone at $\sim 160$~Hz with total duration $\sim 0.03$~s delayed by $\sim 0.1$~s --  could be a fainter 
repetition of the merger.
This would not be in contradiction with e.g. microlensing of the GW primary or GW echoes. Microlensing could explain the observed delay time, corresponding to a fractional travel time of $\sim 2\times10^{-18}$, though it would require characteristics and position of the lens which appear highly unlikely from back-of-the-envelope estimations, further favoring the simpler null hypothesis. 
Our result also does not support echoes with respect to the null hypothesis, in agreement with previous targeted searches analyzing GW151226 and specifically aimed at echoes (see e.g. \cite{abedi2017echoes,nielsen2018parameter,ashton2016comments,westerweck2018low}).

\subsection{other GWTC-1 GWs}

The standard cWB analysis of all other GWTC-1 events does not show any significant deviation from the null hypothesis in the post-coalescence tests, as reported in Table \ref{tab1}.
However, our standard setting constrains signal clustering to a time-window smaller than $\sim 0.3$~s, which is not sufficient to capture the echoes-signal claimed for GW150914 and GW170817 by \cite{abedi2017echoes, abedi2018echoes}.
Motivated by those studies, we decided to expand up to 5~s the on-source time-window allowed for signal clustering in the cWB analysis.

In the case of GW150914, this extension results in the appearance of an onsource post-merger with null coherent SNR and non zero incoherent energy, therefore we can conclude that there is no support for any signal coming from a  direction consistent with GW150914.  

In the case of GW170817, this extension shows a short burst of $\sim 0.01$~s duration and frequency $\sim 130$~Hz at a delay of $\sim 1$~ s with coherent and incoherent components. 
By repeating the same experiment off-source, the p-value of the coherent on source SNR in the post-coalescence is about 20\%, which confirms the null hypothesis.  
The morphology of the cWB feature is compatible with the first harmonic of the signal claimed by \cite{abedi2018echoes}, however, when compared with the widest class of possible post-merger events, as we do here, its p-value becomes uninteresting. 

These results are insufficient to either invalidate or confirm the echoes model of \cite{abedi2018echoes}: in fact, to be able to compare quantitatively the sensitivity of this analysis with the template searches for echoes, 
we would need to calibrate the detection efficiency of the cWB analysis for that specific signal class, which is beyond the scope of the present paper.\\

\section{Conclusions}

The CBC waveforms used in PyCBC, GstLAL and other similar pipelines \cite{GWOSC}, are the result of sophisticated calculations with a well-defined set of physical assumptions. The methodology discussed in this work is aimed at complementing the template-based pipelines and detecting additional features that may show up in the observed time-frequency data as localized excesses of coherent SNR across the network. Indeed, any significant excess coherent energy can indicate the presence of astrophysical structures that reshape the theoretical waveform or point to some still unobserved physics. 

Our analysis of the events in the GWTC-1 catalog \cite{GWTC-1} identifies distinct features after coalescence in the events GW151012 and GW151226 with a  statistical significance which is insufficient to make any claim about their astrophysical origin, but that calls for further investigations of similar event structures in the current and in future observing runs. In this work, we studied the morphology of these features to search for plausible explanations, alternative to the null hypothesis. We look forward to applying this analysis to the larger set of upcoming compact binary coalescences which will be detected in the current observing run of the LIGO-Virgo Collaboration.

In the case of GW151012, the secondary feature is consistent with a subthreshold candidate of a compact binary coalescence. Both its apparent chirpmass and apparent sky direction are different from those of the primary CBC GW signal, which seems to indicate that the secondary is unrelated to the primary and may plausibly be due to the accidental coincidence of a subthreshold CBC candidate with the primary event. 
If this interpretation is correct, the higher sensitivity of the interferometers in the O3 observing run should lead to several more such observations. 

In the case of GW151226, the morphology of the feature that appears after coalescence may suggest a post-merger phenomenon related to the detected GW, also in view of the near superposition of sky localization probability distributions. Again, we interpret this as a simple noise fluctuation. However, it will be interesting to repeat the analysis on the larger set of GWs from the current observation run and check how frequently this type of outliers occurs.

We conclude by remarking that we have established a new general methodology in the framework of cWB, that can be extended to search for a larger variety of additional features, not currently included in the CBC models. In the future, we plan to widen our statistical tests to other regions of the time-frequency representation and to appraise the sensitivity of the analysis for selected models of astrophysical phenomena. 



\medskip

\appendix

\section{cWB sky localizations for GW events within GWTC$-$1 \label{A:skymaps}}

\begin{figure*}[h]
\subfloat[GW150914 HL]{
\includegraphics[clip,width=0.32\textwidth]{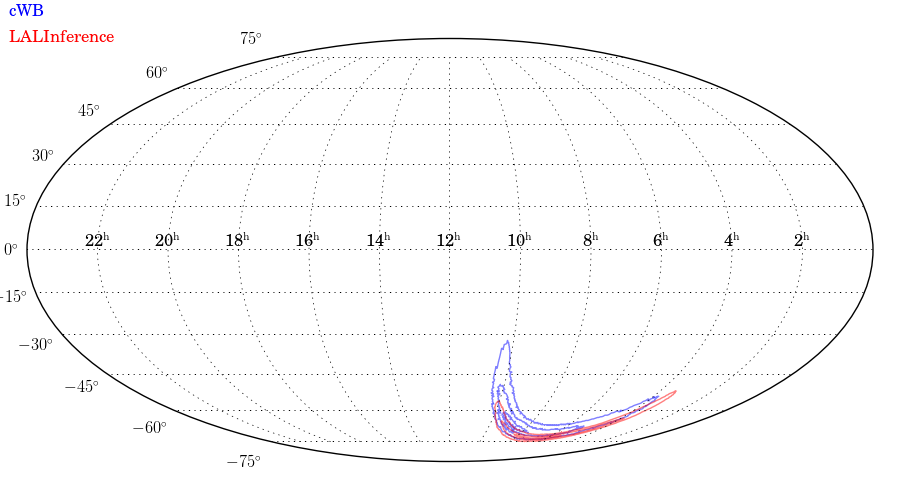}
}
\hfil
\subfloat[GW151012 HL]{
\includegraphics[clip,width=0.32\textwidth]{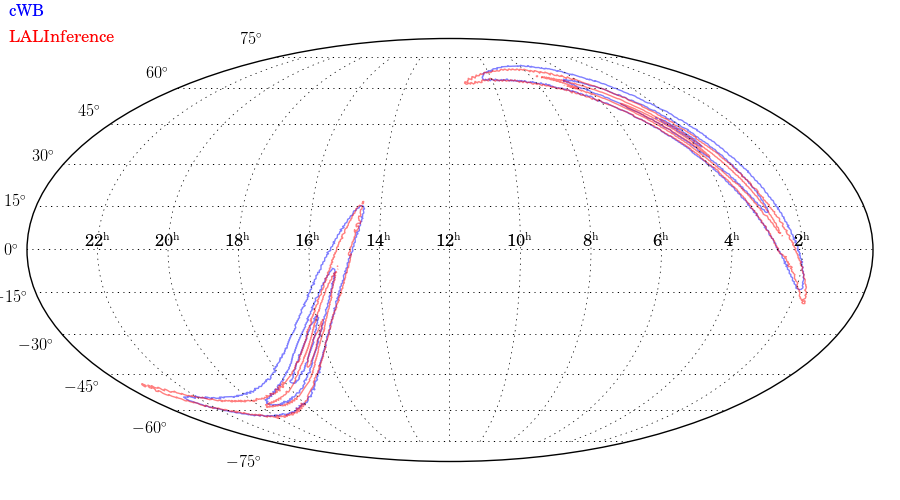}
}
\hfil
\subfloat[GW151226 HL ]{
\includegraphics[clip,width=0.32\textwidth]{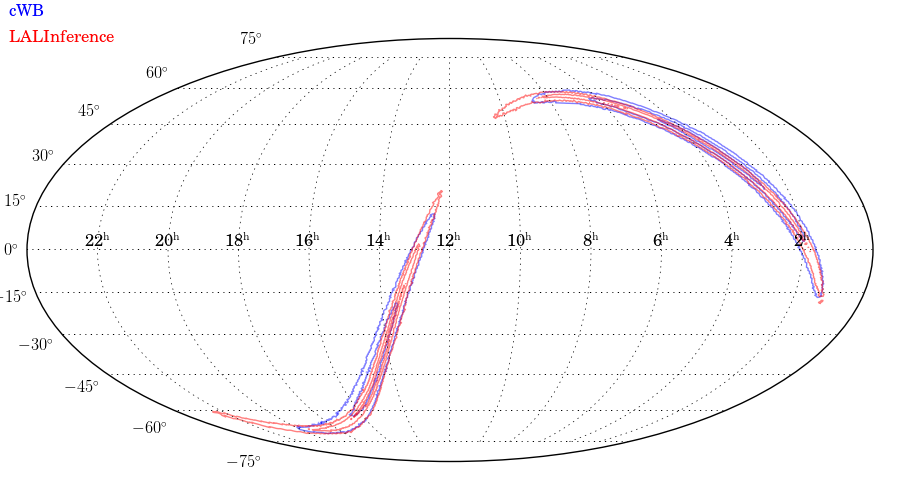}
}

\subfloat[GW170104]{
\includegraphics[clip,width=0.32\textwidth]{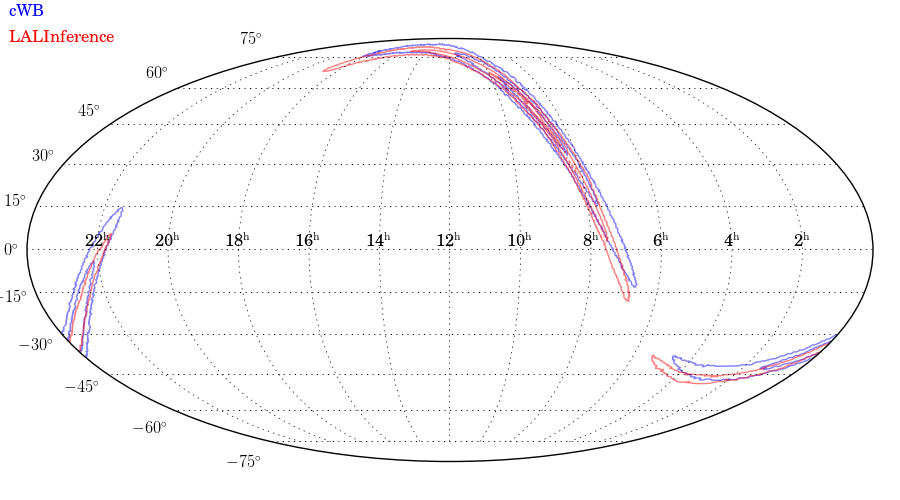}
}
\hfil
\subfloat[GW170608 HL]{
\includegraphics[clip,width=0.32\textwidth]{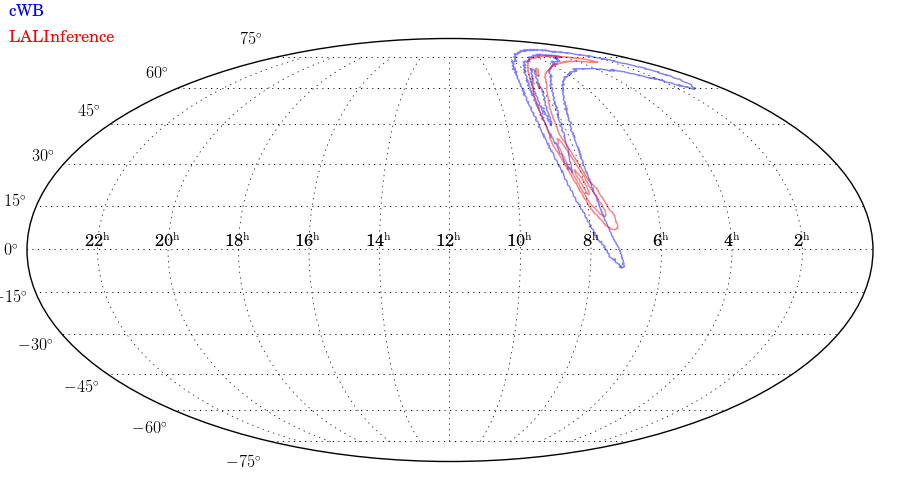}
}
\hfil
\subfloat[GW170729 HLV]{
\includegraphics[clip,width=0.32\textwidth]{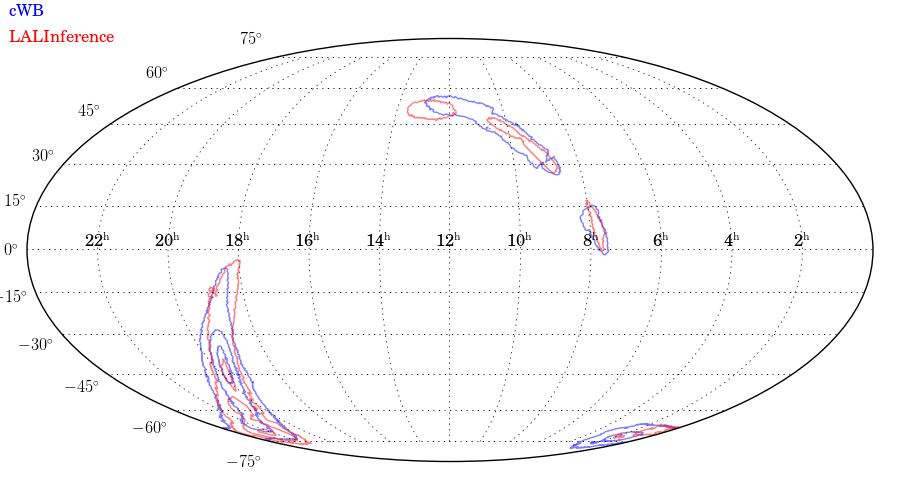}
}

\subfloat[GW170809 HLV]{
\includegraphics[clip,width=0.32\textwidth]{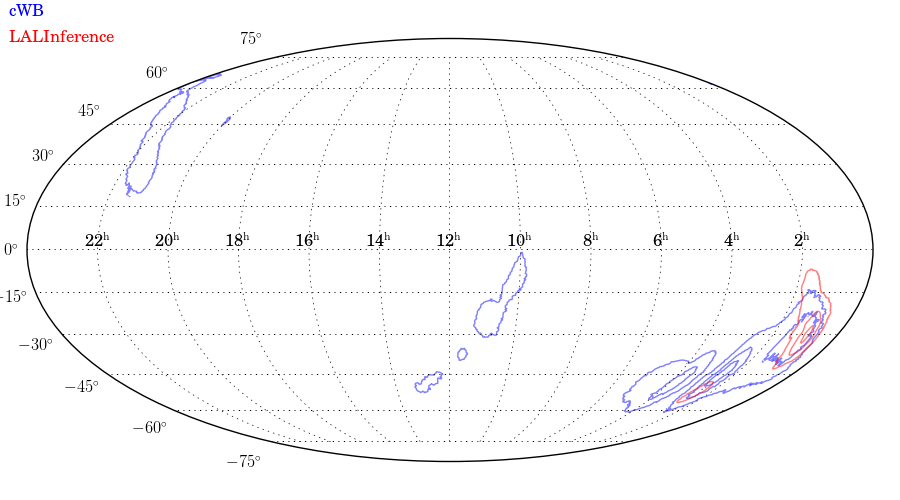}
}
\hfil
\subfloat[GW170814 HLV]{
\includegraphics[clip,width=0.32\textwidth]{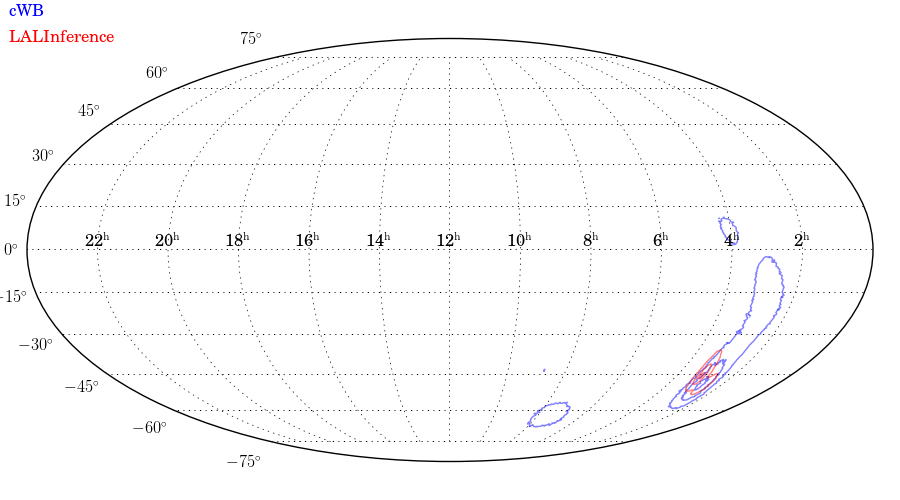}
}
\hfil
\subfloat[GW170817 HLV]{
\includegraphics[clip,width=0.32\textwidth]{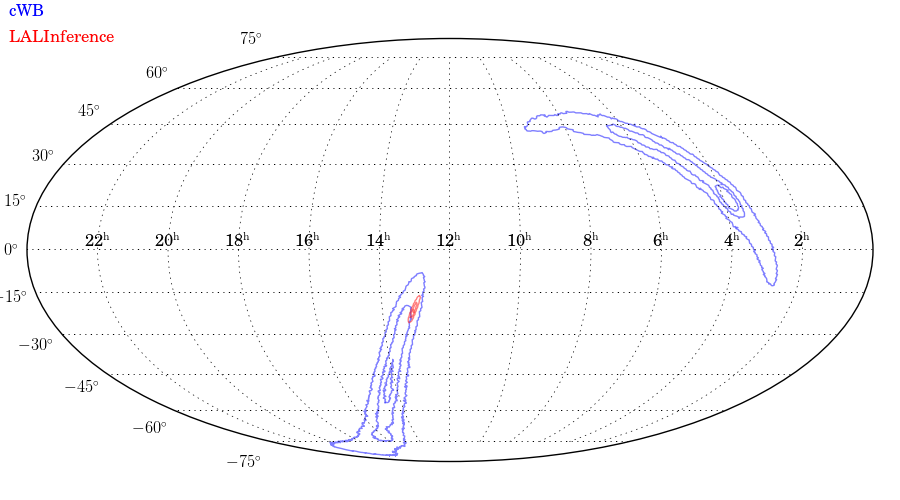}
}

\subfloat[GW170818 HLV]{
\includegraphics[clip,width=0.32\textwidth]{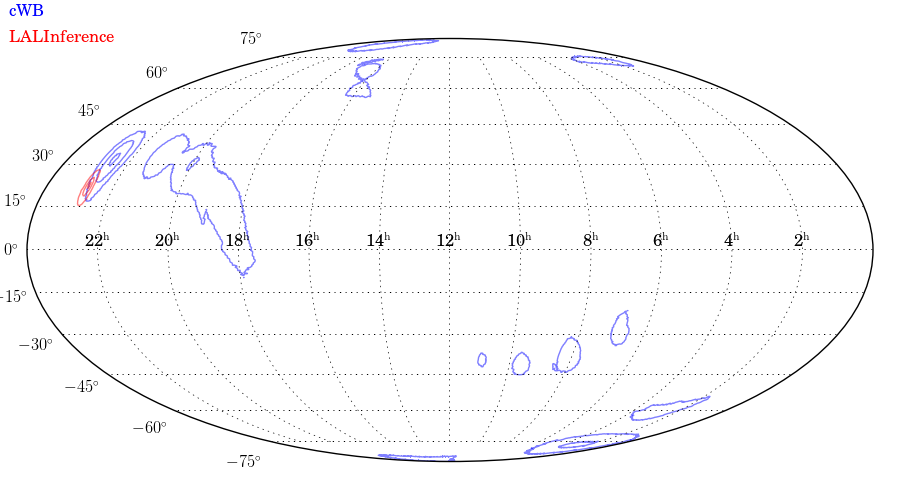}
}
\hfil
\subfloat[GW170823 HL]{
\includegraphics[clip,width=0.32\textwidth]{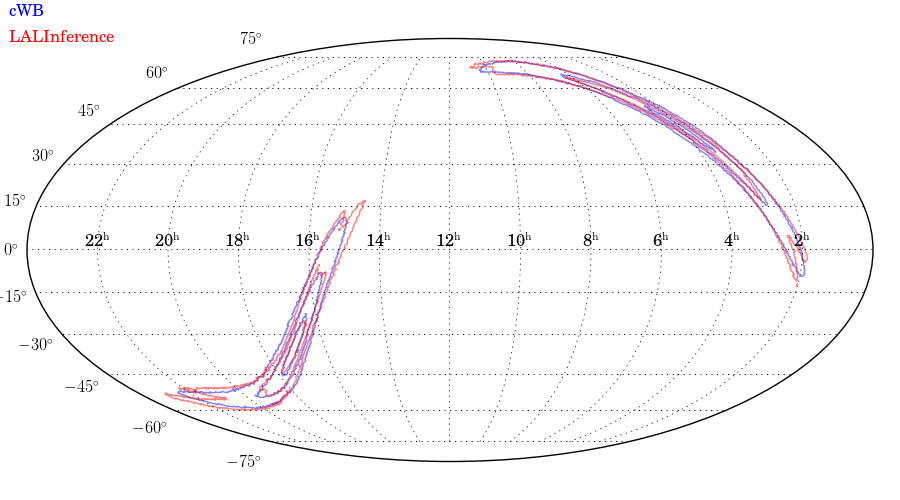}
}


\caption{\label{Mollweide} Mollweide projection of $10\%$, $50\%$ and $90\%$ credible regions for the sky locations of all GW events in GWTC-1 as estimated by cWB and LALInference, in equatorial coordinates. For 5 GW events, i.e. GW170729, GW170809, GW170814, GW170817 and GW170818, the localizations can benefit from the data from a third site, i.e. Virgo.  All the figures have been produced using the code in \cite{reedessick}.
 }
\end{figure*}

The reconstructed sky locations of the gravitational wave events are reported in Figure 
\ref{Mollweide}. Plots show the 10\%, 50\% and 90\% credible regions obtained from cWB (blue) and
and LALInference (green)\footnote{Sky localization probability maps release page for GWTC-1, \url{https://dcc.ligo.org/LIGO-P1800381/public}}. For 5 events the inclusion of the Virgo detector naturally reduces the overall area
of the credible regions for both algorithms. 
In Table \ref{tab:skymaps}, we report the areas of the 50\% and 90\% credible regions for cWB and LALInference, 
together with the extension of the overlapping area between the two estimates.
The last two columns report the nominal probability estimated
by the two algorithms when we restrict the integration inside this overlap area.
Generally the results show that there is a dominant overlap when only two detectors are considered, while
the modeled reconstruction has considerably reduced overlap with respect to the unmodeled one when Virgo is added in
the network.

\medskip

\begin{table*}[htbp]
 \centering
 \begin{tabular}{l c c c c c c c c} 
  \toprule
    {\specialcellbold{GW event }}                 &
    {\specialcellbold{source }}                 &
    {\specialcellbold{network }}                 &
    {\specialcellbold{CR}}                 &
    \multicolumn{3}{c}{Sky area [degrees$^2$]}           &
    \multicolumn{2}{c}{P(Overlap)}      \\
    \cmidrule(lr){5-7}
    \cmidrule(lr){8-9}
    
         &        &       &     &   cWB          &        CBC          & Overlap   &       cWB     &    CBC         \\  
\midrule
GW150914 & BBH & HL & $50\%$ &	135  &	52 & 5   & 0.013 & 0.042 \\
         &      & & $90\%$ &	447  &   180 & 111 & 0.230 & 0.627 \\
GW151012 & BBH & HL & $50\%$ &	628  &	436 & 308   & 0.245 & 0.380 \\
         &      & & $90\%$ &	2196 &   1555 & 1470  & 0.682 & 0.882 \\
GW151226 & BBH & HL &  $50\%$ &	420  &	342 & 185   & 0.218 & 0.267 \\
         &      & & $90\%$ &	1394 &   1033 & 902   & 0.659 & 0.841 \\
GW170104 & BBH & HL &  $50\%$ &	 437 & 205    & 144   & 0.191	& 0.370 \\
         &      & & $90\%$ &	1399 &	924 & 712   & 0.612	& 0.783 \\
GW170608 & BBH & HL &  $50\%$ &	286  &	100 &  23 & 0.035	& 0.087  \\
         &      & & $90\%$ &	1067 &	396&  346 & 0.482 & 0.818  \\
GW170729 & BBH & HLV &  $50\%$ &	297  &	 143& 83  & 0.142	& 0.231 \\
         &      & & $90\%$ &	1477 &   1033&  862	& 0.703	& 0.842  \\
GW170809 & BBH & HLV &  $50\%$ & 299	&	71  &	 52	& 0.082	& 0.390   \\
         &      & & $90\%$ &1748	&	340  &	 250	& 0.268	& 0.793   \\
GW170814 & BBH & HLV &  $50\%$ & 77	&	16  &	 6	& 0.029	& 0.180    \\
         &      & & $90\%$ &	672 &	87  &	 78	& 0.378	&  0.859   \\
GW170817 & BNS & HLV &$50\%$ &	503&	5  &	 0.4	& ---	& 0.036    \\
         &      & & $90\%$ &	1812&	16  &	 16	& 0.007	& 0.902    \\ 
GW170818 & BBH & HLV &  $50\%$ &95.9	& 11 &	--- & ---  & ---   \\
         &      & & $90\%$ &	1707.8&	39  &	 19	& 0.008	& 0.440    \\ 
GW170823 & BBH & HL &  $50\%$ &533	&	430  &	 402	& 0.399	& 0.477    \\
         &      & & $90\%$ & 1635	&	1651  &	 1412	 & 0.847	& 0.853    \\ 
\bottomrule
\end{tabular}
\caption{\label{tab:skymaps} For each of the GW events, we report a comparison between cWB and LALInference (CBC) reconstructed sky localizations: $50\%$ and $90\%$ sky areas with their ovelap are reported. In the last two columns, we give the nominal probability as estimated by cWB and LALInference integrated over the overlap area. For $5$ GW events, i.e. GW170729, GW170809, GW170814, GW170817 and GW170818, the localizations  benefit from the data from a third site, the Virgo interferometer.}
\end{table*}

\section{Posteriors injection simulation \label{A:MTC}}
The  Monte Carlo methodology described here allows for the injection of waveform models from the  public parameter estimation samples released for GWTC-1 events. The samples have been injected every 150~s in the O1-O2 data around each GW event, corresponding to roughly  five days of interferometer coincident time including the GW trigger, with the aim of achieving at least one thousand reconstructions for each GW event. 
For the BBH systems this procedure employs the IMRPhenomPv2 waveform family \cite{PEIMR1,PEIMR2,Schmidt:2014iyl}, a phenomenological waveform family that models signal from the inspiral, merger, ringdown phase taking in to account spin effects, and including simple precession, whereas effects of subdominant (non-quadrupole) modes are not included.
For the BNS system the model IMRPhenomPv2\textunderscore NRTidal  is used instead \cite{NRTidal1}, \cite{NRTidal2}; this model includes numerical relativity (NR)-tuned tidal effects, as well as the spin-induced quadrupole moment.

The cWB unmodeled algorithm is applied to recover the injections, using the same setting of parameters used for O2 analysis, including the veto due to the quality of the  data. All the point estimated reconstructions are used to build empirical distribution of reconstructed signals. This methodology allows to obtain a marginalization over the posterior distribution of the models and to estimate a possible bias due to non Gaussian noise fluctuations in a time span that covers to the selected event.

\bigskip

\begin{acknowledgments}
The authors would like to thank Ik Siong Heng for his constructive inputs.
We also acknowledge useful discussions with Jahed Abedi, Marco Cavaglia, Tom Dent, Sudarshan Ghonge and Tyson Littenberg. 
This research has made use of data, software and/or web tools obtained from the Gravitational Wave Open Science Center (https://www.gw-openscience.org), a service of LIGO Laboratory, the LIGO Scientific Collaboration and the Virgo Collaboration. 
The work by S.Tiwari was supported by SNSF grant $ \#  200020\_182047$.
Finally, we gratefully acknowledge the support of the
State of Niedersachsen/Germany and NSF for provision of computational resources.
\end{acknowledgments}

\bibliography{references}

\end{document}